\documentclass[preprint,aps]{revtex4}
\usepackage{graphicx}
\usepackage{amsmath,amssymb,amsfonts}
\usepackage{natbib}

\newcommand{\be}{\begin{equation}}
\newcommand{\ee}{\end{equation}}

\begin{document}

\title{Second order classical perturbation theory for atom surface scattering: analysis of
asymmetry in the angular distribution}
\author{Eli Pollak}
\affiliation{Chemical Physics Department, Weizmann Institute of Science, 76100 Rehovot,
Israel}
\email{eli.pollak@weizmann.ac.il}
\author{Salvador Miret-Art{\'e}s}
\affiliation{Instituto de Fisica Fundamental, Consejo Superior de
Investigaciones Cientificas, Serrano 123, 28006, Madrid, Spain}
\email{s.miret@iff.csic.es}
\date{\today }

\begin{abstract}
A second order classical perturbation theory is developed and applied to elastic atom corrugated surface scattering. The 
resulting theory accounts for experimentally observed
asymmetry in the final angular distributions. These include qualitative features, such as reduction of the asymmetry with increased incidence energy as well as asymmetry in the location of the rainbow peaks with respect to the specular scattering angle.  The theory is especially applicable to "soft" corrugated
potentials. Analytic expressions for the angular distribution are
derived for the exponential repulsive and Morse potential models. The theory is implemented numerically to a simplified model of the scattering of an Ar atom from a LiF(100) surface.
\end{abstract}

\maketitle


\newpage \renewcommand{\theequation}{1.\arabic{equation}} %
\setcounter{section}{0} \setcounter{equation}{0}

\section{Introduction}

The scattering of atoms from surfaces has been measured extensively during
the past fifty years \cite{oman68,lorenzen68,smith69,smith70}. The measured
angular distribution for heavy atoms, when quantum diffraction effects may
be neglected, is characterized by a few prominent, qualitative features.
Rainbows appear at sub-specular and super-specular angles in the form of
bell shaped maxima in the angular distribution \cite%
{mcclure69,mcclure70,mcclure72}. Typically, the intensity of the
sub-specular peak is larger than that of the super-specular peak \cite%
{rieder85,amirav87,kondo06}. Other features include a reduction of the
angular distance between the rainbow peaks as the incident energy of the
atom is increased \cite{schweizer91,kondo06} or as the angle of incidence
(measured with respect to the vertical) increases \cite{amirav87,schweizer91}%
. Recent reviews of rainbow scattering from surfaces may be found in Refs.
\cite{kleyn91,miret12}.

An early model which gave a qualitative explanation for the asymmetry in the
angular distribution was that of a hard wall corrugated potential \cite%
{steele73,garibaldi75,green78,klein79}. This model may be solved
analytically. It provides a simple explanation for the rainbows - one
readily finds that they originate from the inflection points of the
corrugation. Perhaps more subtle but of not less interest is that the hard
wall corrugated model also provides an explanation for the asymmetry in the
angular distribution. One finds that the potential which faces the incoming
particle leads to the sub-specular peak \cite{klein79}. Its intensity is
higher just as the intensity of the rain that one feels is larger when one
runs into the rain direction rather than away from it.

The hard wall model also provides a partial explanation for the energy
dependence of the distance between the rainbow angles. If one adds a shallow
attractive square well which precedes the wall {\cite%
{steele73,green78,klein79}, one finds that the distance between
the rainbow peaks decreases as the incident energy increases. At
low incident energies, one "feels" the shallow well and due to its
refractive effect on the straight line trajectories, it increases
the distance between the rainbow angles. As the energy is
increased, the refraction decreases and one reaches the repulsive
hard wall limit, in which the rainbow angles are energy
independent. The energy dependence of the angular distance between
the rainbows is thus a sensitive measure of the characteristics of
the physisorption well. }

However, the hard wall class of models is deficient in a number of respects.
For example, the distance between the rainbow angles is independent of the
angle of incidence. The rainbow angles are symmetrically spaced about the
specular angle and this is not always so \cite{amirav87,kondo06}. Moreover,
it is not very realistic, since the interaction between atoms and surfaces
really is rather "soft" especially when dealing with rare gas incident atoms.

With this in mind, we have developed in recent years a classical theory of
atom surface scattering which is based on a perturbation theory in which the
corrugation height is considered to be the small parameter \cite%
{miret12,pollak08,pollak09,pollak10a,pollak10b}. In previous work, we
developed this theory using a perturbation expansion which is valid to first
order in the corrugation height. The first order theory correctly predicts
the incident energy and incident angle dependence of the angular
distribution. The distance between the rainbow angles decreases when the
incoming atom has some time to traverse in the horizontal direction, smearing
out the effect of the corrugation. Therefore, when the angle of incidence is
large, the horizontal velocity is relatively fast and the distance between
the rainbow angles is small. Similarly, at a fixed angle of incidence,
increasing the energy implies also an increase of the horizontal momentum
and this leads to a smaller angular distance between the rainbows.

Similar to the hard wall model, if the potential is purely repulsive, then
the rainbow angles are energy independent, the energy dependence arises only
when the potential includes a physisorption well \cite{miret12}. As in the
hard wall model, the rainbows are symmetrically placed about the specular
angle. The first order perturbation theory does not account for the
asymmetry in the angular scattering unless one imposes an asymmetric
corrugation potential \cite{pollak09}.

The topic of this paper is to show how a second order perturbation
theory, applied to elastic atom-surface scattering, accounts for the asymmetry in the angular
distribution. In principle, the second order theory calls for the
solution of a second order in time equation of motion which is
characterized by a time dependent harmonic frequency and external
force. Such an equation is rather difficult to solve analytically
except in special cases. However, if one uses the fact that energy
is conserved during the scattering, one may replace the second
order in time equation of motion with a first order in time
equation which is readily solved analytically. Using this strategy
we derive in Section II the second order perturbation theory expression for
the final momenta of the particle.

In Section III we use these results to derive the angular distribution expanded up to
second order in the corrugation height. We find that the second order
contribution leads to the correct asymmetry in the angular distribution.
Sub-specular final angles have higher intensity than super-specular angles, due to
the same qualitative effect already understood from the hard wall model.
The perturbation theory also provides an incident angle and energy
dependence for the asymmetry, which decreases as the energy is increased or
as the angle of incidence increases with respect to the vertical direction.
Moreover, the theory accounts for asymmetry in the location of the
rainbow angles with respect to the specular scattering angle.

In Section IV we apply the theory to an exponentially repulsive
potential and a Morse potential model with a sinusoidal
corrugation function. Analytic expressions are derived for the
angular distributions in both cases. As in the first order theory,
the purely repulsive potential gives an energy independent angular
distribution. The energy dependence is directly related to the
physisorption well, which is of course included in the Morse
potential model. Some numerical examples are provided for a simplified model of the
scattering of an Ar atom from a LiF(100) surface. We end with a discussion of the
results, noting for example that one may also observe asymmetry in
the opposite direction, that is that the super-specular
scattering angles are more probable than the sub-specular ones \cite%
{schweizer91}. We also speculate that the second order perturbation
expansion should be useful within a semiclassical description of the
scattering.

\newpage \renewcommand{\theequation}{2.\arabic{equation}} %
\setcounter{section}{1} \setcounter{equation}{0}

\section{Perturbation theory for the momenta}

In this paper, we limit ourselves to a model of in-plane scattering from a
frozen surface. Generalization to the full three dimensional dynamics, as
well as inclusion of surface phonons is straightforward but leads to
somewhat more complicated expressions \cite{miret12}. We thus assume that
the scattering event takes place in the vertical ($z$) and horizontal ($x$)
configuration space. A "standard" model used for the description of the
scattering is based on the assumption that the potential of interaction
depends on the instantaneous distance from the surface and so has the
generic form $V\left( x,z\right) =V\left( z-h\left( x\right) \right) $ where
$h\left( x\right) $ is the small periodic (with lattice length $l$)
corrugation height of the surface. The potential vanishes when the particle
is sufficiently distant from the surface. This potential is then expanded to
first order in the corrugation height, that is:
\begin{equation}
V\left( z,x\right) =V(z)-V^{\prime }(z)h\left( x\right) .  \label{2.1}
\end{equation}%
In principle, since we will develop a second order perturbation theory with
respect to the corrugation, one should expand to include also the second
order term, however, this does not create any fundamental differences in the
results, only leads to a more complicated algebra, so that we will remain
with the standard first order expansion for the potential.

We will study the
classical scattering for a particle with mass $M$ and vertical and
horizontal momenta $p_{z}$ and $p_{x}$ respectively. The Hamiltonian
governing the motion is thus:%
\begin{equation}
H=\frac{p_{x}^{2}+p_{z}^{2}}{2M}+V\left( z,x\right) .  \label{2.2}
\end{equation}
The exact equations of motion governing the vertical and horizontal
distances are%
\begin{equation}
M\ddot{z}_{t}+V^{\prime }(z_{t})-V^{\prime \prime }(z_{t})h(x_{t})=0
\label{2.3}
\end{equation}
\begin{equation}
M\ddot{x}_{t}-V^{\prime }(z_{t})h^{\prime }\left( x_{t}\right) =0.
\label{2.4}
\end{equation}
where the primes denote derivatives with respect to the argument and
the dots, time differentiation. The particle is assumed to be initiated at
the time $-t_{0}$ with initial vertical (negative) momentum $p_{z_{i}}$ and
(positive) horizontal momentum $p_{x_{i}}$. The zero-th order motion
(expansion to order $h^{0}$) is decoupled, the vertical motion is governed
by the vertical Hamiltonian%
\begin{equation}
H_{z0}=\frac{p_{z}^{2}}{2M}+V(z)  \label{2.5}
\end{equation}%
and the horizontal motion is that of a free particle
\begin{equation}
H_{x0}=\frac{p_{xi}^{2}}{2M}.  \label{2.6}
\end{equation}%
with constant velocity $v_{x}=p_{xi}/M$. In the zero-th order
motion, the particle impacts the surface at time $t=0$ and then
leaves the interaction region by the time $t_{0}$ which is taken
to be sufficiently large to assure
that the scattering event is over. Finally, we take the limit of $%
t_{0}\rightarrow \infty $.

We then expand the horizontal and vertical motions to second order in the
corrugation height:
\begin{eqnarray}
x_{t} &=&x_{0,t}+x_{1,t}+x_{2,t}+O\left( h\left( x\right) ^{3}\right)
\label{2.7} \\
p_{x,t} &=&p_{x0,t}+p_{x1,t}+p_{x2,t}+O\left( h\left( x\right) ^{3}\right)
\label{2.8} \\
z_{t} &=&z_{0,t}+z_{1,t}+z_{2,t}+O\left( h\left( x\right) ^{3}\right)
\label{2.9} \\
p_{z,t} &=&p_{z0,t}+p_{z1,t}+p_{z2,t}+O\left( h\left( x\right) ^{3}\right) .
\label{2.10}
\end{eqnarray}%
The zero-th order vertical solution obeys the equation of motion:%
\begin{equation}
M\ddot{z}_{0,t}+V^{\prime }(z_{0,t})=0  \label{2.11}
\end{equation}%
and the first order correction to the vertical motion obeys the equation of
motion:%
\begin{equation}
M\ddot{z}_{1,t}+V^{\prime \prime }(z_{0,t})z_{1,t}-V^{\prime \prime
}(z_{0,t})h(x_{0,t})=0.  \label{2.12}
\end{equation}

In the horizontal direction the motion is to zero-th order that of a free
particle (parallel momentum conservation) such that:%
\begin{equation}
x_{0,t}=x_{0,-t_{0}}+\frac{p_{xi}}{M}\left( t+t_{0}\right) \equiv x_{0,0}+%
\frac{p_{xi}}{M}t.  \label{2.13}
\end{equation}%
The Jacobian of the transformation between the initial value of the
horizontal coordinate and its value upon impact $x_{0,0}$, is unity. The
first order correction to the horizontal motion is determined by
\begin{equation}
M\ddot{x}_{1,t}-V^{\prime }(z_{0,t})h^{\prime }(x_{0,t})=0  \label{2.14}
\end{equation}%
while the second order equation of motion for the horizontal coordinate is:%
\begin{equation}
M\ddot{x}_{2,t}-\bar{V}^{\prime }(z_{0,t})h^{\prime \prime }\left(
x_{0,t}\right) x_{1,t}-\bar{V}^{\prime \prime }(z_{0,t})z_{1,t}h^{\prime
}\left( x_{0,t}\right) =0.  \label{2.15}
\end{equation}
One then readily finds that the horizontal momenta up to second order are
expressed in terms of the lower order solutions as:
\begin{equation}
p_{x1,t}=\int_{-t_{0}}^{t}dt\bar{V}^{\prime }(z_{0,t})h^{\prime }\left(
x_{0,t}\right)  \label{2.16}
\end{equation}%
and
\begin{equation}
p_{x2,t}=\int_{-t_{0}}^{t}dt\left[ \bar{V}^{\prime }(z_{0,t})h^{\prime
\prime }\left( x_{0,t}\right) x_{1,t}+\bar{V}^{\prime \prime
}(z_{0,t})z_{1,t}h^{\prime }\left( x_{0,t}\right) \right] .  \label{2.17}
\end{equation}

The corrugation function $h\left( x\right) $ is periodic with period $l$. In
the following, we will employ the simplest possible periodic corrugation
function%
\begin{equation}
h\left( x\right) =h\sin \left( \frac{2\pi x}{l}\right)  \label{2.18}
\end{equation}%
but here too, we note that it is straightforward, only increasingly complex
to use a higher harmonic expansion for the corrugation function \cite%
{miret12,pollak09}. Using the symmetry of the motion along the vertical
direction ($V^{\prime }(z_{0,t})$ is symmetric with respect to the time) we
find that the first order contribution to the final horizontal momentum is
\cite{pollak08}:%

\begin{equation}
p_{x1,f}=p_{z_{i}}K\cos \left( \frac{2\pi x_{0,0}}{l}\right)  \label{2.19}
\end{equation}%
with
\begin{equation}
K=\frac{2\pi h}{lp_{z_{i}}}\int_{-\infty }^{\infty }dtV^{\prime
}(z_{0,t})\cos \left( \omega _{x}t\right) ,  \label{2.20}
\end{equation}
the horizontal frequency is defined as:%
\begin{equation}
\omega _{x}=\frac{2\pi }{l}\frac{p_{xi}}{M}  \label{2.21}
\end{equation}%
and we have taken the limit of $t_{0}\rightarrow \infty $. The dimensionless
quantity $K$ is termed the rainbow shift angle, as will also become evident
from the expression for the angular distribution, as shown in the next
Section. This first order contribution to the horizontal coordinate is then
seen to take the form \cite{pollak08}:%
\begin{equation}
x_{1,t}=\cos \left( \frac{2\pi x_{0,0}}{l}\right) \int_{-\infty
}^{t}dt^{\prime }F_{c}\left( t^{\prime }\right) -\sin \left( \frac{2\pi
x_{0,0}}{l}\right) \int_{-\infty }^{t}dt^{\prime }F_{s}\left( t^{\prime
}\right)  \label{2.22}
\end{equation}%
where%
\begin{eqnarray}
F_{c}\left( t\right) &=&h\frac{2\pi }{lM}\int_{-\infty }^{t}dt^{\prime }\bar{%
V}^{\prime }(z_{0,t^{\prime }})\cos \left( \omega _{x}t^{\prime }\right) ,
\label{2.23} \\
F_{s}\left( t\right) &=&h\frac{2\pi }{lM}\int_{-\infty }^{t}dt^{\prime }\bar{%
V}^{\prime }(z_{0,t^{\prime }})\sin \left( \omega _{x}t^{\prime }\right) .
\label{2.24}
\end{eqnarray}

The first order contribution to the vertical momentum necessitates the
solution of Eq. \ref{2.12}. This equation is equivalent in form to that of a
forced oscillator with a time dependent frequency, which is rather difficult
to solve analytically. We note however that to first order in the
corrugation height, energy conservation implies that for any time
\begin{equation}
0=\frac{p_{xi}p_{x1,t}+p_{z0,t}p_{z1,t}}{M}+V^{\prime
}(z_{0,t})z_{1,t}-V^{\prime }(z_{0,t})h(x_{0,t}).  \label{2.25}
\end{equation}%
This relation then determines the first order contribution to the vertical
motion $p_{z1,t}$ in terms of the first order contribution to the horizontal
motion:
\begin{equation}
p_{z1,f}=\frac{p_{xi}}{p_{zi}}p_{x1,f}.  \label{2.26}
\end{equation}

Not less important is to note that the first order energy conservation
relation \ref{2.25} also leads to a first order in time equation of motion
for the first order contribution to the vertical motion:
\begin{equation}
\dot{z}_{1,t}=\frac{\dot{p}_{z0,t}}{p_{z0,t}}z_{1,t}+\frac{1}{p_{z0,t}}%
\int_{-t_{0}}^{t}dth\left( x_{0,t}\right) \frac{d}{dt}\bar{V}^{\prime
}(z_{0,t})  \label{2.27}
\end{equation}%
whose solution is readily written as:
\begin{equation}
z_{1,t}=p_{z0,t}\int_{-t_{0}}^{t}dt^{\prime }\frac{1}{p_{z0,t^{\prime }}^{2}}%
\int_{-t_{0}}^{t^{\prime }}dt^{\prime \prime }h(x_{0,t^{\prime \prime }})%
\frac{d}{dt^{\prime \prime }}\bar{V}^{\prime }(z_{0,t^{\prime \prime }}).
\label{2.28}
\end{equation}%
By taking the time derivative of Eq. \ref{2.28}, one may see that this is a
specific solution to the second order in time equation of motion \ref{2.12}
for the first order correction. It is this observation that
allows us to obtain closed form expressions for the final momenta, up to
second order. Using the specific sinusoidal form for the corrugation (Eq. \ref{2.18})
Eq. \ref{2.28} may be rewritten as
\begin{equation}
z_{1,t}=\sin \left( \frac{2\pi x_{0,0}}{l}\right)
p_{z0,t}\int_{-t_{0}}^{t}dt^{\prime }G_{c}\left( t^{\prime }\right) +\cos
\left( \frac{2\pi x_{0,0}}{l}\right) p_{z0,t}\int_{-t_{0}}^{t}dt^{\prime
}G_{s}\left( t^{\prime }\right)  \label{2.29}
\end{equation}%
where we used the notation:%
\begin{eqnarray}
G_{c}\left( t\right) &=&\frac{h}{p_{z0,t}^{2}}\int_{-\infty }^{t}dt^{\prime }%
\frac{d\bar{V}^{\prime }(z_{0,t^{\prime }})}{dt^{\prime }}\cos \left( \omega
_{x}t^{\prime }\right) ,  \label{2.30} \\
G_{s}\left( t\right) &=&\frac{h}{p_{z0,t}^{2}}\int_{-\infty }^{t}dt^{\prime }%
\frac{d\bar{V}^{\prime }(z_{0,t^{\prime }})}{dt^{\prime }}\sin \left( \omega
_{x}t^{\prime }\right) .  \label{2.31}
\end{eqnarray}

With these preliminaries, using Eqs. \ref{2.17}, \ref{2.22} and \ref{2.29}
we find after some manipulation that the second order contribution to the
final horizontal momentum simplifies to:%
\begin{equation}
p_{x2,f}\equiv p_{xi}K_{cc}  \label{2.32}
\end{equation}%
with
\begin{eqnarray}
K_{cc} &=&\frac{2\pi Mh}{lp_{xi}}\int_{-\infty }^{\infty }dt\left[ \cos
\left( \omega _{x}t\right) \frac{d\bar{V}^{\prime }(z_{0,t})}{dt}%
\int_{-\infty }^{t}dt^{\prime }G_{s}\left( t^{\prime }\right) -\frac{2\pi }{%
lM}\bar{V}^{\prime }(z_{0,t})\sin \left( \omega _{x}t\right) \int_{-\infty
}^{t}dt^{\prime }F_{c}\left( t^{\prime }\right) \right] .  \notag \\
&&  \label{2.33}
\end{eqnarray}
\ \ The second order contribution to the final vertical momentum is then
found through energy conservation to be:
\begin{equation}
p_{z2,f}=\frac{p_{xi}}{p_{zi}}p_{x2,f}+\frac{p_{x1,f}^{2}+p_{z1,f}^{2}}{%
2p_{zi}}.  \label{2.34}
\end{equation}

Eqs. \ref{2.32}-\ref{2.34} are the central results of this Section. An
explicit solution for the second order contribution to the final momenta and
its dependence on the point of impact on the surface, has been derived. In
principle, one can follow the same methodology to obtain all order
contributions to the final momenta, however the complexity increases
accordingly. One may also put in higher harmonics into the corrugation
function, however, as we shall see in the next Section, the second order
perturbation theory suffices to provide a qualitative explanation for the
observed angular distributions, their asymmetry and rainbow structure, and
angle of incidence and energy dependence.

\newpage \renewcommand{\theequation}{3.\arabic{equation}} %
\setcounter{section}{2} \setcounter{equation}{0}

\section{Perturbation theory for the final angular distribution}

The (negative) angle of incidence with respect to the vertical vector to the
surface is by definition%
\begin{equation}
\theta _{i}=\tan ^{-1}\left( \frac{p_{xi}}{p_{zi}}\right) .  \label{3.1}
\end{equation}%
The final angular distribution is (with $\bar{x}=\frac{2\pi x_{0,0}}{l}$):
\begin{equation}
P\left( \theta _{f}\right) =\frac{1}{2\pi }\int_{0}^{2\pi }d\bar{x}\delta
\left( \theta _{f}-\tan ^{-1}\left( \frac{p_{x_{f}}\left( \bar{x}\right) }{%
p_{z_{f}}\left( \bar{x}\right) }\right) \right)   \label{3.2}
\end{equation}%
where $p_{xf}\left( \bar{x}\right) $ and $p_{zf}\left( \bar{x}\right) $ are
the final momenta as determined from Hamilton's equations of motion. Using
the second order expansion results from the previous section, this may be
rewritten as:%
\begin{equation}
P\left( \theta _{f}\right) \simeq \frac{1}{2\pi \cos ^{2}\left( \theta
_{f}\right) }\int_{0}^{2\pi }d\bar{x}\delta \left( \tan \theta _{f}+f\left(
\bar{x}\right) \right)   \label{3.3}
\end{equation}%
with%
\begin{eqnarray}
f\left( x\right)  &=&\frac{p_{xi}+p_{x1,f}+p_{x2,f}}{p_{zi}-p_{z1,f}-p_{z2,f}%
}  \notag \\
&=&\frac{K\cos \left( \bar{x}\right) +\tan \theta _{i}\left( 1+K_{cc}\right)
}{1-\tan \theta _{i}K\cos \left( \bar{x}\right) -K_{cc}\tan ^{2}\theta _{i}-%
\frac{K^{2}\cos ^{2}\left( \bar{x}\right) }{2\cos ^{2}\theta _{i}}}.
\label{3.4}
\end{eqnarray}

Using the notation $y=K\cos \bar{x}$ we note that imposing that the argument
of the Dirac "delta" function in Eq. \ref{3.3} vanishes leads to a quadratic
equation in $y$ whose solutions are denoted as $y^{\ast }$:%
\begin{equation}
y_{\pm }^{\ast }=\frac{1\pm \sqrt{1+2\left[ \tan \left( \theta _{i}+\theta
_{f}\right) -\tan \theta _{i}\right] \left[ \tan \left( \theta _{i}+\theta
_{f}\right) +K_{cc}\tan \theta _{i}\right] }}{\tan \left( \theta _{f}+\theta
_{i}\right) -\tan \theta _{i}}  \label{3.5}
\end{equation}%
The condition that $-K\leq y^{\ast }\leq K$ and noting that the perturbation theory implies that the magnitudes of the rainbow shift angle 
should be small, typically such that  $\left\vert
K\right\vert <1$, implies that only the minus sign solution is physical. To
gain some further insight, and staying consistent within the second order
perturbation theory we may expand this solution:
\begin{equation}
y^{\ast }=y_{1}^{\ast }+y_{2}^{\ast }+O\left( h^{3}\right)   \label{3.6}
\end{equation}%
where the subscript denotes first and second order in $\left( \theta
_{f}+\theta _{i}\right) $ respectively, to find:
\begin{eqnarray}
y_{1}^{\ast } &=&-\tan \left( \theta _{f}+\theta _{i}\right) ,  \label{3.7}
\\
y_{2}^{\ast } &=&-\tan \theta _{i}\left( K_{cc}+\frac{\tan ^{2}\left( \theta
_{f}+\theta _{i}\right) }{2}\right) .  \label{3.8}
\end{eqnarray}%

To obtain the angular distribution it is necessary to determine the
derivative of the deflection function at the points $y^{\ast }$ ,which may
be written as:%
\begin{equation}
\frac{df}{d\bar{x}}=-K\sin \bar{x}\frac{df}{dy}.  \label{3.9}
\end{equation}%
Rainbows are found when the derivative of the deflection function vanishes
that is when $f^{\prime }\left( \bar{x}\right) =0$, provided that $\frac{df}{dy}$
does not diverge at these points. One finds that:
\begin{equation}
\frac{df}{dy}=\frac{1+\frac{y^{2}}{2}+y\tan \theta _{i}\left(
1+K_{cc}\right) }{\cos ^{2}\theta _{i}\left( 1-\tan \theta _{i}y-K_{cc}\tan
^{2}\theta _{i}-\frac{y^{2}}{2\cos ^{2}\theta _{i}}\right) ^{2}}
\label{3.10}
\end{equation}%
so that the rainbows occur either when
\begin{equation}
\sin \bar{x}=0  \label{3.11}
\end{equation}%
or $\frac{df}{dy}=0$:%
\begin{equation}
1+y_{R}\tan \theta _{i}\left( 1+K_{cc}\right) +\frac{y_{R}^{2}}{2}=0
\label{3.12}
\end{equation}%
where the subscript reminds us that this is the condition for the rainbow
angles. For the condition $\sin \left(\bar{x}\right) =0$ we have two solutions, $%
\bar{x}=0$ or $\bar{x}=\pi $. In the second case, we have a quadratic equation with
solutions%
\begin{equation}
y_{R,\pm }=-\tan \theta _{i}\left( 1+K_{cc}\right) \pm \sqrt{\tan ^{2}\theta
_{i}\left( 1+K_{cc}\right) ^{2}-2}.  \label{3.13}
\end{equation}%
These solutions will only be valid if
\begin{equation}
\tan ^{2}\theta _{i}\left( 1+K_{cc}\right) ^{2}\geq 2  \label{3.14}
\end{equation}%
and this implies collisions which are close to grazing collisions, or more
specifically, $\left\vert \theta _{i}\right\vert \gtrsim 60^{o}$.

The angular distribution is then%
\begin{equation}
P\left( \theta _{f}\right) =\frac{\cos ^{2}\theta _{i}\left( 1-\tan \theta
_{i}y^{\ast }-\tan ^{2}\theta _{i}K_{cc}-\frac{y^{\ast ^{2}}}{2\cos
^{2}\theta _{i}}\right) ^{2}}{\pi \cos ^{2}\theta _{f}\left\vert 1+y^{\ast
}\tan \theta _{i}\left( 1+K_{cc}\right) +\frac{y^{\ast ^{2}}}{2}\right\vert
\sqrt{K^{2}-y^{\ast ^{2}}}}.  \label{3.15}
\end{equation}%
To make further sense of this deceptively simple but in fact quite
complicated expression we resort to perturbation theory. Expanding up to
second order using the second order expansion of $y^{\ast }$ as in Eqs. \ref%
{3.7} and \ref{3.8} the angular distribution simplifies to

\begin{equation}
P\left( \theta _{f}\right) \simeq \frac{ 1+\tan \theta _{i}\tan
\left( \theta _{f}+\theta _{i}\right) +K_{cc}\tan ^{2}\theta _{i}+\frac{%
3\tan ^{2}\left( \theta _{f}+\theta _{i}\right) }{2}\left( \tan ^{2}\theta
_{i}-1\right)  }{\pi \cos ^{2}\left( \theta _{f}+\theta
_{i}\right) \sqrt{K^{2}-y^{\ast ^{2}}}}.  \label{3.16}
\end{equation}%
Rainbows occur whenever $y^{\ast }=\pm K$, but the strength of the
divergence depends on the sign of $y^{\ast }$ due to the linear term in the
numerator. This gives rise to the asymmetry of the distribution. The rainbow
condition implies to second order that%
\begin{equation}
\tan \left( \theta _{f}+\theta _{i}\right) =\pm K-\tan \theta _{i}\left(
K_{cc}+\frac{K^{2}}{2}\right)  \label{3.17}
\end{equation}
showing that the location of the rainbows is no longer symmetric about the
specular angle. Using the minus sign for $\pm \vert K\vert$ corresponds
to a rainbow angle which is\ \textit{smaller }than the specular angle. Since
the linear term in the intensity goes as $\tan \theta _{i}\tan \left( \theta
_{f}+\theta _{i}\right) $ and $\theta _{i}$ is negative, we get that $\tan
\theta _{i}\tan \left( \theta _{f}+\theta _{i}\right) >0$ and the amplitude
for the smaller scattering angle is increased relative to the amplitude for
the larger scattering angle. If, as is typically the case, the rainbow shift angle $K>0$, this implies (through the relation $y=Kcos(\bar{x})$) that 
$\bar{x}=0$. When considering the
corrugation potential $-V^{\prime }\left( z\right) h\sin \left( \bar{x}\right) $
this implies that when the particle hits the part of the potential which
points towards the particle, the scattering intensity is higher. This is
also qualitatively consistent with rainbow scattering from a hard wall
corrugated potential.

It is also worthwhile noting that when limiting the derivation to only first
order perturbation theory (denoted by the subscript) then the angular
distribution takes the symmetric form \cite{miret12,pollak08}:%
\begin{equation}
P\left( \theta _{f}\right) =\frac{1}{\pi \cos ^{2}\left( \theta _{f}+\theta
_{i}\right) \sqrt{K^{2}-\tan ^{2}\left( \theta _{f}+\theta _{i}\right) }}
\label{3.18}
\end{equation}%
demonstrating clearly that the asymmetry observed in the angular
distribution is a second order effect. Finally it is instructive to compare
the second order result Eq. \ref{3.16} with the angular distribution found
for scattering from a hard wall potential with the same sine corrugation
function as in Eq. \ref{2.18}:
\begin{equation}
P_{hw}\left( \theta _{f}\right) =\frac{\left( 1+\tan \left( \frac{\theta
_{f}+\theta _{i}}{2}\right) \tan \left( \theta _{i}\right) \right) }{\pi
\cos ^{2}\left( \frac{\theta _{f}+\theta _{i}}{2}\right) \sqrt{%
K_{hw}^{2}-4\tan ^{2}\left( \frac{\theta _{f}+\theta _{i}}{2}\right) }}
\label{3.19}
\end{equation}%
where
\begin{equation}
K_{hw}=\frac{4\pi h}{l}.  \label{3.20}
\end{equation}

The hard wall potential exhibits rainbow angles that are symmetrically
placed around the specular angle. The asymmetry in the distribution which
comes from the second term in the numerator is half as large as the
asymmetric part in Eq. \ref{3.16}. Here one notes though that the hard wall
model is qualitatively different from the potential used in the present
model. In the hard wall model, the location of the hard wall varies with the
corrugation while the hard wall limit of the model we have been using (Eq. %
\ref{2.1}) leads to a hard wall whose location is independent of the
horizontal coordinate. However, the qualitative feature of a larger
scattering amplitude originating from the wall facing the incident particle
is the same in both cases. \bigskip

\newpage \renewcommand{\theequation}{4.\arabic{equation}} %
\setcounter{section}{3} \setcounter{equation}{0}

\section{Analytical models and numerical results}

\subsection{Repulsive exponential potential}

Perhaps the simplest soft potential is obtained by replacing the hard wall with a purely repulsive
exponential potential%
\begin{equation}
\bar{V}_{e}\left( z\right) =V_{e}\exp \left( -\alpha z\right)  \label{4.1}
\end{equation}%
where the inverse length $\alpha $ is the stiffness parameter of the
exponential potential. $V_{e}$ has the dimensions of energy and expresses
the "strength" of the exponential interaction. The trajectory for the
vertical motion at energy $E_{z}$ is known analytically \cite{miret12}:%
\begin{equation}
\exp \left( \alpha z_{t}\right) =\frac{V_{e}}{2E_{z}}\left[ 1+\cosh \left(
\Omega t\right) \right] ,\text{ \ }  \label{4.2}
\end{equation}%
with
\begin{equation}
\Omega ^{2}=\frac{2\alpha ^{2}E_{z}}{M} \label{4.3}
\end{equation}%
and $E_z$ is the incident energy in the vertical direction. The rainbow angle shift (see Eq. \ref{2.20}) for this model is:
\begin{equation}
K_{e}=K_{hw}\frac{\pi \bar{\Omega}}{\sinh \left( \pi \bar{\Omega}\right) }
\label{4.4}
\end{equation}%
with%
\begin{equation}
\bar{\Omega}=\frac{\omega _{x}}{\Omega }=\frac{2\pi }{\alpha l}\left\vert
\tan \theta _{i}\right\vert  \label{4.5}
\end{equation}%
and in the hard wall limit, that is when the stiffness parameter $\alpha
\rightarrow \infty $, $K_{e}\rightarrow K_{hw}$.

It is a matter of some algebra, based on the integral (for further
manipulations see Section IV of Ref. \cite{pollak09})
\begin{equation}
\int_{-\infty }^{\infty }dt\frac{\cos \left( \bar{\Omega}t\right)
}{\left[ \cos \Phi +\cosh \left( t\right) \right]
}=\frac{2\pi\sinh
\left( \Phi \bar{\Omega}\right) }{\sin \Phi \sinh \left( \pi \bar{\Omega}%
\right) }  \label{4.6}
\end{equation}%
and the identity:%
\begin{equation}
\frac{\left[ 1+\cosh \left( t\right) \right] ^{2}}{\sinh ^{2}\left( t\right)
}=\frac{d}{dt}\left( t-\frac{2\cosh t+2}{\sinh t}\right)  \label{4.7}
\end{equation}%
to find that:%
\begin{equation}
K_{cc,e}=\frac{K_{e}^{2}}{\tan ^{2}\theta _{i}}\left( 1-\frac{\pi \bar{\Omega%
}\cosh \left( \bar{\Omega}\pi \right) }{\cos ^{2}\theta _{i}\sinh \left(
\bar{\Omega}\pi \right) }\right) .  \label{4.8}
\end{equation}

This implies, that for the repulsive exponential potential, also to second
order in perturbation theory, the angular distribution is independent of the
incident energy. It only depends on the angle of incidence. The angular
distribution, to second order in the perturbation theory is given by Eq. \ref%
{3.16}. In the hard wall limit one has that the second order parameter $%
K_{cc,e}\rightarrow -K_{hw}^{2}$.

\subsection{Morse potential model}

The dependence of the angular distribution on the incident energy
comes from the existence of a shallow physisorbed well in the
potential of interaction of the atom with the surface. This is
well modelled by the Morse potential
\begin{equation}
V_{M}\left( z\right) =V_{0}\left[ \left( \exp \left( -\alpha z\right)
-1\right) ^{2}-1\right]  \label{4.9}
\end{equation}%
which has a physisorption well depth $V_{0}$. In this case, the rainbow
shift parameter (Eq. \ref{2.20}) is%
\begin{equation}
K_{M}=K_{hw}\frac{\pi \bar{\Omega}\cosh \left( \bar{\Omega}\Phi \right) }{%
\sinh \left( \pi \bar{\Omega}\right) }\text{ \ }  \label{4.10}
\end{equation}%
with
\begin{equation}
\cos \Phi =-\sqrt{\frac{V_{0}}{E_{z}+V_{0}}}.  \label{4.11}
\end{equation}%
and $\bar{\Omega}$ as defined in Eq. \ref{4.5}. For a fixed angle of
incidence, the angle $\Phi $ decreases from $\pi $ to $\pi /2$ as the energy
is increased, causing the rainbow shift parameter to decrease accordingly.
The attractive well leads to a narrowing of the distance between the rainbow
angles, as the incident energy is increased.

The trajectory for the Morse potential at an incident energy $E_{z}$ is
known analytically \cite{miret12}:
\begin{equation}
\exp \left( \alpha z_{t}\right) =-\frac{\cos \Phi }{\sin ^{2}\Phi }\left[
\cosh \left( \Omega t\right) +\cos \Phi \right] ,  \label{4.12}
\end{equation}%
with the frequency $\Omega $ as given in Eq. \ref{4.3}. We then use the same
integrals as for the exponential repulsive potential and the identity:%
\begin{equation}
\frac{1}{p_{z0,t^{\prime }}^{2}}=\frac{\alpha ^{2}}{M^{2}\Omega ^{2}}\frac{d%
}{d\left( \Omega t\right) }\left( \frac{\Omega t\sinh \left( \Omega t\right)
-\cosh \left( \Omega t\right) \left[ 1+\cos ^{2}\Phi \right] -2\cos \Phi }{%
\sinh \left( \Omega t\right) }\right)  \label{4.14}
\end{equation}%
to find that:
\begin{equation}
K_{cc}=K_{M}^{2}\left( \frac{1}{\tan ^{2}\left\vert \theta _{i}\right\vert }+%
\frac{1}{\sin ^{2}\left\vert \theta _{i}\right\vert }\left( \Phi \bar{\Omega}%
\tanh \left( \bar{\Omega}\Phi \right) -\pi \bar{\Omega}\coth \left( \pi \bar{%
\Omega}\right) \right) -\tanh \left( \bar{\Omega}\Phi \right) \bar{\Omega}%
\cos \Phi \sin \Phi \right) .  \label{4.15}
\end{equation}

In the limit that the stiffness parameter $\alpha \rightarrow \infty $ this
reduces to the previous result, that is $K_{cc}=-K_{hw}^{2}$. In the limit
of a purely repulsive exponential potential ($\Phi \rightarrow 0$) we regain
the exponential model result. In the high energy limit, that is when $\Phi
\rightarrow \pi /2$ we have that%
\begin{equation}
\lim_{E_{z}\rightarrow \infty }K_{cc}=K_{M}^{2}\left( \frac{1}{\tan
^{2}\left\vert \theta _{i}\right\vert }-\frac{\pi }{2}\frac{\bar{\Omega}}{%
\sin ^{2}\left\vert \theta _{i}\right\vert }\coth \left( \frac{\pi \bar{%
\Omega}}{2}\right) \right)  \label{4.16}
\end{equation}%
while in the low energy limit we have that $\Phi \rightarrow \pi $ so that%
\begin{equation}
\lim_{E_{z}\rightarrow 0}K_{cc}=K_{M}^{2}\left( \frac{1}{\tan ^{2}\left\vert
\theta _{i}\right\vert }\right) .  \label{4.17}
\end{equation}%
and this reduces to $-K_{hw}^2$ in the hard wall limit ($\alpha\rightarrow\infty$). We thus find that the second order coefficient is positive at low energies and becomes negative as the energy is increased. Its magnitude reaches a maximum and then decreases with increasing energy. This leads to a reduction of the asymmetry in the angular distribution as the energy is further increased.

\subsection{Numerical examples}

\begin{figure}
\includegraphics* [width=8cm,height=8cm]{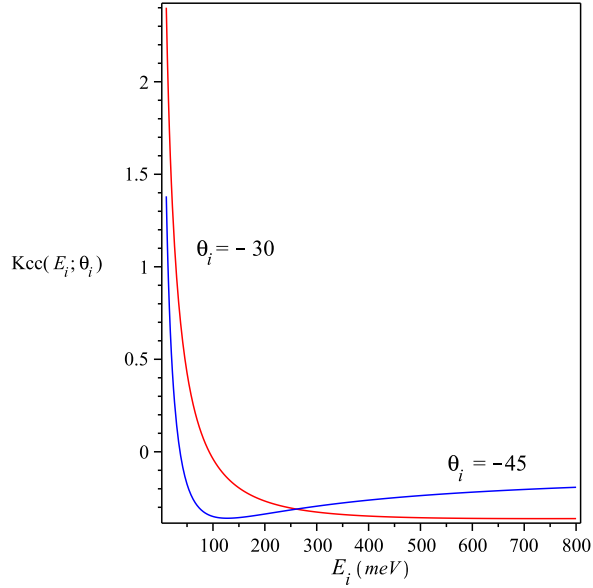}
\caption{ (color online) The second order coefficient $K_{cc}$ for
the Morse potential model given by Eq. \protect\ref {4.15} is
plotted as a function of the incident energy (in meV), for two
incident angles, $-45^o$ and $-30^o$.} \label{fig:one}
\end{figure}

The scattering of an Ar atom from the LiF(100) surface has been studied
in some detail, both experimentally \cite{kondo06} as well as theoretically \cite{miret12}. The
experimental angular distribution, measured with a fixed angle of
90 degrees between the incident beam and the detector showed a
number of distinct qualitative features. The
distance between the rainbow peaks decreased as a function of
increasing incident energy. The asymmetry in the angular
distribution was such that the intensity of the rainbow peak was
higher for final angles which were less than $\pi/4$, however the asymmetry
decreased as the incident energy decreased. All of these features
are accounted for within the present second order perturbation
theory.

\begin{figure}
\includegraphics* [width=8cm,height=8cm]{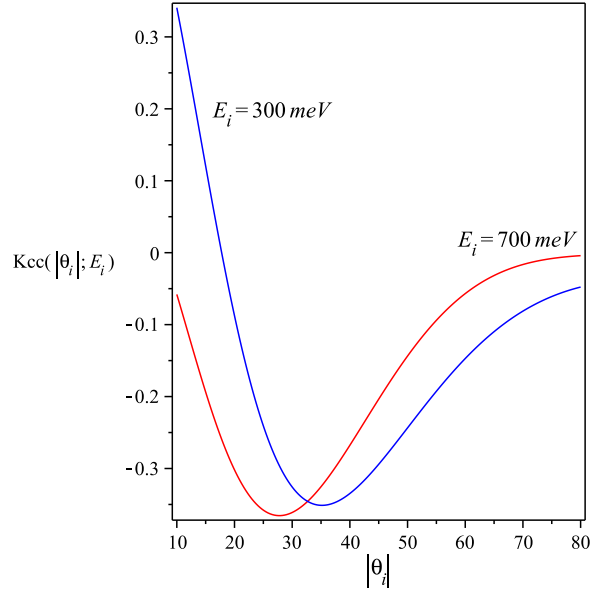}
\caption{ (color online) The second order coefficient $K_{cc}$ for
the Morse potential given by Eq. \protect\ref {4.15} is plotted as
a function of the incident angle for two incident energies, 300
meV and 700 meV.} \label{fig:two}
\end{figure}

To demonstrate this we employ parameters used previously to fit
the experimental results. In particular, the following values have
been used for the Morse potential model \cite{miret12}: $h = 0.25
$ a. u., $l = 4 \, \AA$, $\alpha l = 3$ and $V_0 = 88$ meV.

As has been previously stated, the second order coefficient,
$K_{cc}$, plays a key role in the asymmetry of the angular
distributions. The dependence of this coefficient on the incident
energy and incident angle is shown in Fig.\ref{fig:one}. In the
range of energies (315-705 meV) probed by the experiment \cite{kondo06} the
magnitude of the coefficient is a decreasing function of the
incident energy. At very low energies, the asymmetry of the
angular distribution is expected to be very important since this
second order coefficient becomes relatively large. When the
incident energy is very low, the approaching atom "feels" the
corrugation for a longer period of time thus distinguishing
between the case that the atom approaches the downhill or uphill
part of the corrugation. Given the large magnitude of the coefficient one should expect that in this low energy limit the perturbation theory will
not be accurate.  As the incident energy increases,
$K_{cc}$ becomes small, approaching the purely repulsive model
result, the asymmetry in the angular distribution is reduced and the perturbation theory result should be rather accurate.

The variation of the second order coefficient with respect to the angle of incidence is plotted at two different incident energies in
Fig.\ref{fig:two}. These results have a number of interesting features. Firstly, at low angles of incidence and low energies the coefficient becomes positive. Secondly, the parabolic structure implies that similar
values are obtained at different incident angles, indicating that the asymmetry is not necessarily a monotonic function of the angle of incidence. Fourthly, as may also be discerned from Eq. \ref{4.15} when the angle of incidence tends to $\pi/2$ (grazing angle) the coefficient vanishes.

\begin{figure}
\includegraphics* [width=8cm,height=8cm]{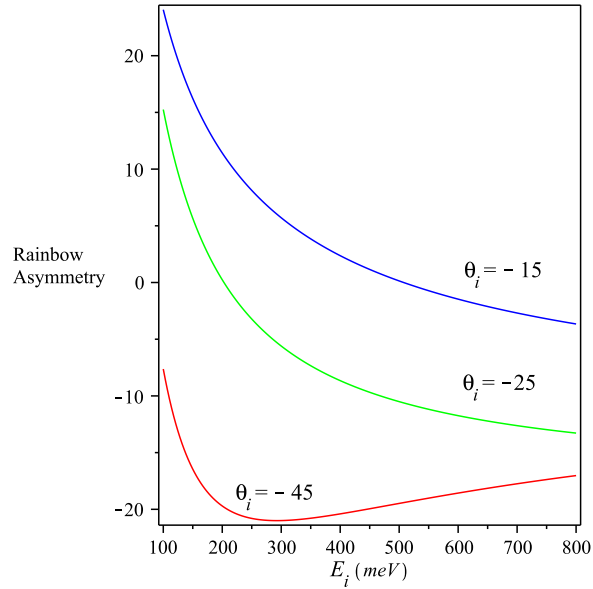}
\caption{ (color online) Rainbow asymmetry evaluated from Eq.
\protect\ref {3.17} for the Morse potential model and plotted as a
function of the incident energy and three incident angles,
-15$^o$, -25$^o$ and -45$^o$ (see text).} \label{fig:three}
\end{figure}

Another property which emerges from the second order perturbation theory is that in contrast to the hard wall model, the location of the rainbows is no longer symmetrically distributed about the specular angle. A measure of this "rainbow asymmetry" is obtained by considering
the difference between the angular distance of the rainbow angles from the specular angle. In the symmetric case, this difference of course vanishes.  In
Fig. \ref{fig:three}, using Eq. \ref{3.17}, we plot the rainbow asymmetry (in degrees) obtained by subtracting the distance of the subspecular rainbow peak from the specular angle from the distance of the superspecular rainbow angle from the specular angle.   From this figure we note that, depending on the angle of incidence, the rainbow asymmetry can change sign. In addition the dependence of the asymmetry on the incidence energy is not necessarily monotonic.

\begin{figure}
\includegraphics* [width=8cm,height=8cm]{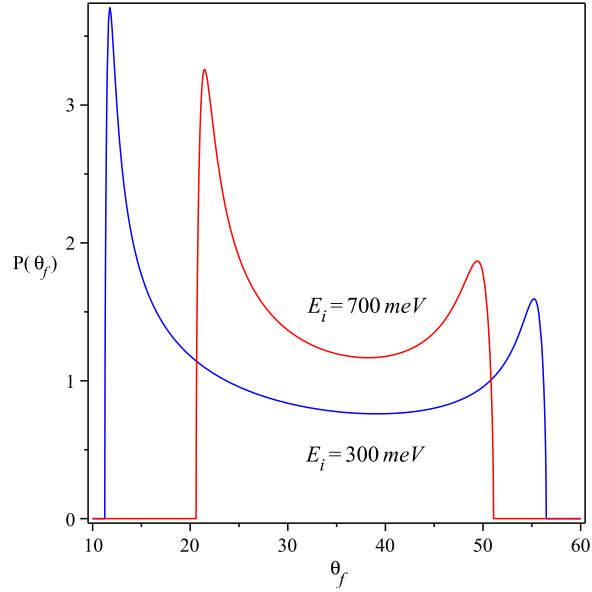}
\caption{ (color online) The final angular distribution as given
by Eq. \protect\ref {3.15} is plotted for two incident energies,
300 meV and 700 meV (covering the experimental range) and for an
incident angle of - 45$^o$.} \label{fig:four}
\end{figure}

\begin{figure}
\includegraphics* [width=8cm,height=8cm]{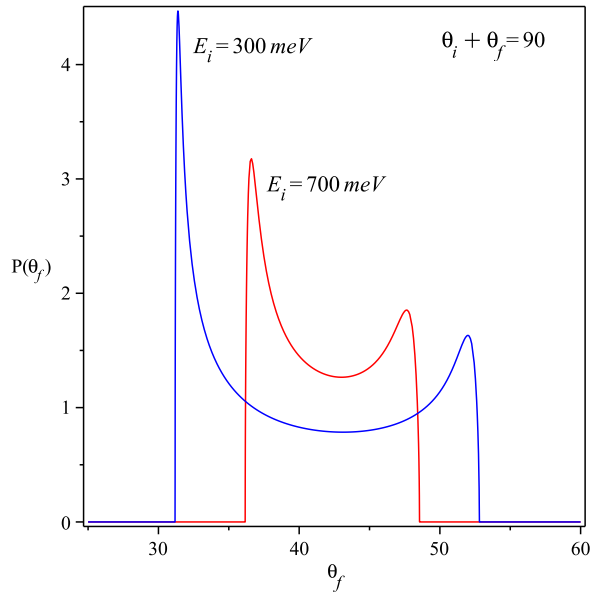}
\caption{ (color online) The final angular distributions as given
by Eq. \protect\ref {3.15}, but with a fixed angle of $\pi/2$
between incident and outgoing beams is plotted for two incident
energies, 300 meV and 700 meV, covering the experimental range.}
\label{fig:five}
\end{figure}

When considering the angular distribution, one distinguishes
between two different experiments. In one class, the angle of
incidence is kept fixed and the detector is moved to measure the
final angular dependence of the outcoming flux. In a different
(easier) experimental setup, as used in the measurements of Ref.
\cite{kondo06}, the angle between the incident and final beam is
kept fixed and only the angle of incidence is varied. The results
for the angular distribution for the former case are shown in Fig.
\ref{fig:four} for two different incidence energies. The
singularities associated with the rainbow peaks have been smoothed
by approximating the step function with a hyperbolic tangent
function. As the hyperbolic tangent function tends to the step
function the rainbow peaks become higher but it becomes more
difficult to resolve their magnitude numerically. The present depiction
suffices to show the asymmetry in the distribution. One notes that
the asymmetry decreases significantly as the energy is increased
from 300 meV to 700 meV. The fixed ($\pi/2$) angular distribution,
plotted as a function of the exit angle
($\theta_f=\pi/2+\theta_i$) is shown in Fig. \ref{fig:five}.
Qualitatively, the results are similar to those shown in Fig.
\ref{fig:four}, however the distance between the rainbow peaks
becomes smaller.

\bigskip

\newpage \renewcommand{\theequation}{5.\arabic{equation}} %
\setcounter{section}{4} \setcounter{equation}{0}

\section{Discussion}

A second order perturbation theory with respect to the corrugation
height has been developed for the elastic scattering of atoms from
a periodic corrugated surface. The second order theory correctly
accounts for experimentally observed asymmetry in the measured
angular distributions. Expressions have been derived for the
energy and incident angle dependence of the angular distributions
and their asymmetry. In contrast to the hard wall models, the
second order theory provides also the dependence of the asymmetry
on both angle of incidence and energy of the particle. Analytical
expressions for this dependence were derived for a purely
repulsive exponential model potential as well as a Morse potential
which exhibits the characteristic physisorption well felt by the
incoming atom. Numerical results were shown for parameter values
which fit qualitatively the scattering of Ar from a LiF(100) surface.
In contrast to the corrugated hard wall model, the second order
theory also accounts for asymmetry in the location of the rainbow
angles.

In this context it should be noted though that the asymmetry is not always such that the subspecular rainbow peak is the preferred one. The opposite is found for the scattering of Ar from an H covered Tungsten surface \cite{schweizer91}. This indicates for example that the elastic theory presented in this paper, may not always be sufficient. For example, phonon friction which is larger in the vertical direction than in the horizontal direction will tend to shift the final angular distribution towards superspecular peaks. The three dimensional structure of the surface can also affect the asymmetry in the distribution \cite{miret12}.

As already noted, 
the present theory was limited to elastic scattering. It can be
further expanded to include the interaction of the particle with
surface phonons. One should expect that a first order theory with
respect to the coupling to the surface phonons should suffice, at
least when considering the angular distribution. It is though
possible but rather cumbersome to follow the methodology presented
in this paper to also treat the coupling to the phonons to second
order in the coupling constant.

The classical first order perturbation theory was used in previous
work \cite{miret12,pollak11} to study the sticking of atoms
scattered from surfaces. Here too, one could employ the present
second order theory to study sticking. It would be of interest to
understand how much the asymmetry will change the sticking
probabilities.

The first order perturbation theory fails especially for grazing
angles, where the change in the horizontal momentum can no longer
be considered as small with respect to the magnitude of the
incident horizontal momentum.  The second order perturbation
theory should improve the theory but the extent is not clear.
Detailed comparison with numerically exact classical mechanics
simulations of the scattering would be helpful in this respect.

Finally, we note that the first order perturbation theory has been
used extensively within a semiclassical context
\cite{hubbard84,daon12}. It should be of interest to see whether
the present second order perturbation theory can be employed
semiclassically, so that also the resulting semiclassical
diffraction patterns will exhibit the correct asymmetry.

\bigskip

\textbf{Acknowledgment} This work has been supported by grants of
the Israel Science Foundation, the German-Israel Foundation for
basic research and the Einstein center at the Weizmann Institute
of Science and the Ministerio de Economia y Competitividad under
project FIS2011-29596-C02-C01. We also acknowledge support
from the COST Action MP 1006.

\end{document}